\documentclass[10pt,twocolumn,twoside]{IEEEtran}
\usepackage[utf8]{inputenc}
\IEEEoverridecommandlockouts
\usepackage{cite}
\usepackage{mathtools}
\usepackage{adjustbox}
\usepackage{amssymb,amsfonts,amsmath,amsthm,upgreek}
\usepackage{algorithm}
\usepackage{algorithmic}
\usepackage{graphicx,multirow}
\usepackage{textcomp}
\usepackage{color,soul}
\usepackage{enumerate}
\usepackage{pgfplots}
\usepackage{physics}
\usepackage{url}
\usepackage{booktabs,colortbl}
\usepackage{array}
\usepackage{dblfloatfix}
\usepackage{bm}
\usepackage{dcolumn}
\usepackage{longtable}
\usepackage{makecell}
\usepackage{adjustbox}
\usepackage[caption=false, font=footnotesize]{subfig}

\newcommand{\RNum}[1]{\uppercase\expandafter{\romannumeral #1\relax}}
\newcolumntype{d}[1]{D{.}{.}{#1}}
\newcolumntype{L}{>{\raggedright\arraybackslash}m{3.2cm}}
\def\BibTeX{{\rm B\kern-.05em{\sc i\kern-.025em b}\kern-.08em
		T\kern-.1667em\lower.7ex\hbox{E}\kern-.125emX}}
\usepackage{tikz}
\begin{document}
    
    \title{Identifying Secure Operating Ranges for \\DER Control using Bilevel Optimization}
	
\author{Kshitij~Girigoudar,~\IEEEmembership{Member,~IEEE,}          
	and Line~A.~Roald,~\IEEEmembership{Member,~IEEE}
	
	\thanks{The authors are with the Department of Electrical and Computer Engineering, University of Wisconsin-Madison, USA (e-mail: girigoudar@wisc.edu,
roald@wisc.edu).}}
    
    {}

	\maketitle
    \begin{abstract}
Active distribution grids are accommodating an increasing number of controllable electric loads and distributed energy resources (DERs). A majority of these DERs are managed by entities other than the distribution utility, such as individual customers or third-party aggregators, who control the loads and DERs without consideration of any distribution grid constraints. This makes it challenging for a distribution system operator (DSO) to allow third-party aggregators and transmission operators to fully exploit the flexibility offered by these resources while also ensuring that distribution grid constraints such as voltage magnitude limits are not violated. In this paper, we develop a bilevel optimization-based framework to determine the aggregate power flexibility that can be obtained from an unbalanced distribution grid while ensuring that there is no disaggregation solution that leads to grid constraint violations. 
The results are a set of constraints and operating rules that are easy to communicate, and which provide the entities that procure flexibility from DERs (e.g. transmission operators or third-party aggregators) with the ability to freely implement their own disaggregation strategy without intervention from the DSO. The proposed approach is tested on two unbalanced distribution feeders and our simulation results indicate that it is possible to determine a wide range of aggregate power flexibility, as long as a simple set of rules for DER control activation are followed.
\end{abstract}

    \section*{Nomenclature}

\textit{Sets \& Indices}
\begin{IEEEdescription}[\IEEEusemathlabelsep\IEEEsetlabelwidth{${V}_{\text{d}i},{V}_{\text{q}i}\in \mathbb{R}^{3}$}]
\item[$\mathcal{N}$] Set of nodes excluding slack node
\item[${\Phi}$] Set of phases, ${\Phi}=\{a,b,c\}$
\item[$i$] Node index, $i \in \mathcal{N}$
\item[$\phi$] Phase index, $\phi \in \Phi$ 
\end{IEEEdescription}

\textit{Parameters}
\begin{IEEEdescription}[\IEEEusemathlabelsep\IEEEsetlabelwidth{${V}_{\text{d}i},{V}_{\text{q}i}\in \mathbb{R}^{3}$}]
\item[$n$] Number of single-phase nodes, $n=3|\mathcal{N}|$
\item[${v}_{\text{d}i}^\phi,v_{\text{q}i}^\phi$] Rectangular components of current voltage phasor with magnitude $|v_i^\phi|$
\item[$\underline{v},\overline{v}$] Lower and upper voltage magnitude limits
\item[${V}_{\text{d}i},{V}_{\text{q}i}\in \mathbb{R}^{3}$] Rectangular components of current three-phase voltage phasor $V_i$
\item[${p}_{i}^\phi,q_{i}^\phi$] Active and reactive components of current power injection
\item[${P}_{i},Q_{i}\in \mathbb{R}^{3}$] Active and reactive components of current three-phase power injection 
\item[${p}_{\text{G},i}^\phi,q_{\text{G},i}^\phi$] Active and reactive components of current power generation 
\item[${p}_{\text{L},i}^\phi,q_{\text{L},i}^\phi$] Active and reactive components of current power load demand 
\item[$\underline{p}_{\text{G},i}^\phi,\overline{p}_{\text{G},i}^\phi$] Lower and upper limits for active power generation
\item[$\underline{p}_{\text{L},i}^\phi,\overline{p}_{\text{L},i}^\phi$] Lower and upper limits for active power load demand 
\item[$\Delta \underline{p},\Delta \overline{p}$] Lower and upper limits for maximum available active power flexibility
\item[${pf}_{\text{L},i}^\phi$] Load constant power factor setting
\item[$|{s}_{\text{G},i}^\phi|$] Solar PV inverter apparent power capacity
\item[${\gamma}_{\text{G},i}^\phi$] Solar PV inverter constant power ratio setting
\item[${P}_{G},P_{L}\in \mathbb{R}^{n}$] Active power generation and load vectors at current operating point

\end{IEEEdescription}

\textit{Variables}
\begin{IEEEdescription}[\IEEEusemathlabelsep\IEEEsetlabelwidth{$\Delta \boldsymbol{p}_{\mathbf{G},i}^\phi,\Delta \boldsymbol{p}_{\mathbf{L},i}^\phi~$}]
\item[$\Delta \boldsymbol{p}^-,\Delta \boldsymbol{p}^+$] Lower and upper limit variables for aggregate active power flexibility  
\item[$\Delta \boldsymbol{p}_{\mathbf{G},i}^\phi,\Delta \boldsymbol{p}_{\mathbf{L},i}^\phi$] Active power generation and load flexibility variables 
\item[$\Delta \boldsymbol{P}_{\mathbf{G}},\!\Delta \boldsymbol{P}_{\mathbf{L}}\!\in \!\mathbb{R}^{n}$] Vector of active power generation and load flexibility variables 
\item[$\boldsymbol{q}_{\mathbf{G},i}^\phi, \boldsymbol{q}_{\mathbf{L},i}^\phi$] Reactive power generation and load variables 
\item[$ \boldsymbol{Q}_{\mathbf{G}}, \boldsymbol{Q}_{\mathbf{L}}\in \mathbb{R}^{n}$] Vectors of reactive power generation and load variables 
\item[$\boldsymbol{P},\boldsymbol{Q}\in \mathbb{R}^{n}$] Vectors of active and reactive power injection variables
\item[$\boldsymbol{v}_{\mathbf{d}i}^\phi,\boldsymbol{v}_{\mathbf{q}i}^\phi$] Rectangular components of voltage variable with magnitude $|\boldsymbol{v}_i^\phi|$
\item[$\boldsymbol{V}_{\mathbf{d}i},\boldsymbol{V}_{\mathbf{q}i}\in \mathbb{R}^{3}$] Rectangular components of three-phase voltage variable 
\item[$\boldsymbol{V}_{\mathbf{d}},\boldsymbol{V}_{\mathbf{q}}\in \mathbb{R}^{n}$] Vectors of voltage variables in rectangular form
\item[$\boldsymbol{\gamma}_{\mathbf{G},i}^\phi$] Solar PV inverter power ratio variable
\item[$\overline{\boldsymbol{q}}_{\mathbf{G},i}^\phi$] Solar PV inverter maximum available reactive power variable
\item[$\boldsymbol{X}_\textbf{u},\boldsymbol{X}_\textbf{l}$] Vectors of upper- and lower-level variables of bilevel problem
\end{IEEEdescription}

    \section{Introduction}

\IEEEPARstart{T}{oday's} distribution grids are experiencing an increasing amount of smart electric loads and distributed energy resources (DERs), such as electric vehicles, smart home appliances, rooftop solar photovoltaic (PV) systems and energy storage. With growing penetration of dispatchable DERs, distribution grids have evolved from being considered as passive loads to take on a more active role in system operation~\cite{d2009global}. Importantly, dispatchable DERs owned by individual customers can provide flexibility in the form of energy services such as operating reserves to transmission systems. However, each customer-owned DER can only contribute a comparatively small amount of flexibility. To achieve a high enough level of flexibility to act as a reserve, aggregation and coordinated control of many DERs is required. Such aggregation and control services are typically provided by DER aggregators. In this paper, we refer to them as \emph{third-party aggregators} to highlight that they are not part of a distribution utility.  

Typically, a distribution feeder comprises of DERs managed by various entities whose control objectives can be different from each other. The flexibility offered by DERs relies on assumptions about who controls them. DERs managed directly by their owners, who choose not to enroll their DERs with a third-party aggregator, may act as uncontrollable loads and may increase variability rather than offering any flexibility. On the one hand, DERs can be controlled individually by the distribution system operator (DSO), who can use them to manage distribution grid constraints. Furthermore, third-party aggregators can control multiple DERs to achieve an aggregated response across a large population of DERs, typically without awareness of their exact geographical location and without any consideration of the distribution grid constraints. To comply with the recently announced FERC Order 2222~\cite{FERC} in the USA, independent system operators (ISOs) will be required to allow third-party DER aggregators to participate in electricity markets. This may lead to a significant increase in the number of aggregators and the overall potential for DER flexibility. The focus of this paper is to propose a method that can be used to determine limits on the combined DER flexibility that can be offered from within a distribution feeder without violating any distribution grid constraints.   

There are numerous works in literature determining the flexibility of individual DERs~\cite{yi2022operating,nazir2019convex,nazir2021grid} or aggregate power flexibility of distribution grids~\cite{silva2018estimating,xu2020quantification,polymeneas2016aggregate,ageeva2019analysis,heleno2015estimation,tan2020estimating}. Most of these methods assume that network constraints such as voltage magnitude limits are non-binding, and hence do not consider them. Approaches that do consider network constraints include geometric methods, where the aggregate power flexibility is determined using polytopic projection or calculating the Minkowski sum of the polytope sets, as discussed for balanced distribution grids in~\cite{muller2017aggregation,zhao2017geometric,hao2014aggregate} and for unbalanced three-phase network models in~\cite{chen2019aggregate,wang2021aggregate,cui2021network,chen2021leveraging}. However, when assessing the total available range of flexibility, these works assume that the DSO directly controls all DERs, i.e. the DSO acts as an aggregator and can leverage DER resources to both provide transmission grid services and manage distribution grid constraints. Other works model the interactions between aggregators and DSOs~\cite{zhang2016real,zhang2016real1,sheikhahmadi2021bi,riaz2017generic}, or between end consumers and aggregators~\cite{riaz2017generic,zugno2013bilevel}, including methods that provide DSOs with the opportunity to change control signals from the aggregator that would lead to violations\cite{ross2021strategies}. 

Existing methods rely on strong assumptions on the frequency of communication between the DSO, ISO and aggregators or either require or assume an allocation of flexibility among DERs (i.e. which DERs are going to provide the desired flexibility). In a practical setting, where multiple aggregators and DER owners may be competing for the same limited grid capacity, it is problematic that the methods employed to determine the safe amount of DER flexibility may create limits that favor certain DERs over others. This is particularly when the DSO does not have any information about the cost of contracting flexibility from individual DERs. 
To address this issue, we propose a different approach that focuses on assessing a range of aggregate DER flexibility, while making more practical assumptions regarding communication between the DSO, aggregator and ISO. We propose a hierarchical decision making process where an upper-level authority (i.e. DSO) takes decisions based on the unknown, and thus assumed to be worst-case, responses by a lower-level entity (i.e. aggregators). This framework naturally lends itself to modelling as a  bilevel optimization problem.  A key feature of the proposed approach is that it does not prescribe or limit which DERs need to be providing flexibility, as long as the cumulative amount of actuation stays within the allowable range of aggregate DER flexibility. Thus, the method places minimal limits on both how much flexibility should be procured from each of the aggregators active within the feeder (allowing ISOs to procure flexibility from the aggregator that has the most competitive offer), and does not limit how aggregators choose to actuate the DERs that they control to achieve a certain set-point after they have been contracted to provide flexibility. 

The main contributions of the paper are as follows: 1) We propose a bilevel formulation to determine aggregate DER flexibility range that can be offered by distribution grids which can guarantee that \emph{any} disaggregation solution (including the worst-case condition) is feasible, thus enabling more safe disaggregation without further coordination with the DSO. 2) We present a computationally tractable, strong duality based reformulation of the bilevel problem which can be solved efficiently using an iterative approach. 3) The proposed method is tested on two unbalanced distribution feeders, where we investigate how the range of the feeder’s aggregate power flexibility can be maximized by enforcing a set of simple rules for DER actuation.

The rest of the paper is organized as follows:  Section~\ref{sec:prob_desc} describes the problem setup. Section~\ref{sec:prob_for} presents the bilevel problem formulation. Section~\ref{sec:prob_refor} details the analytical reformulation of the problem and an iterative solution approach to solve the problem efficiently. We present the numerical results on two test cases in Section~\ref{sec:sim_res}. Section~\ref{sec:conc} concludes. 
    \section{Problem Setup}
\label{sec:prob_desc}
We next introduce the problem setup by providing a motivating example, and giving an overview of the bilevel optimization framework.

\subsection{Motivating Example}
To comply with the FERC Order 2222~\cite{FERC}, ISOs in the USA will be required to allow DER aggregators to  participate in electricity markets. While the clearing of electricity markets by ISOs currently only considers transmission grid constraints, increased levels of generation and reserve provision from DERs located at the distribution grid level may require that ISOs include distribution grid constraints in their optimal dispatch algorithms. Here, we present a method that enables DSOs to identify limits on how much DER flexibility the ISO can procure from a given feeder without causing constraint violations. We demonstrate the intended use of this method with a small, illustrative example. 

\begin{figure}[!t]
	\centering	        	
	\includegraphics[width=0.39\textwidth]{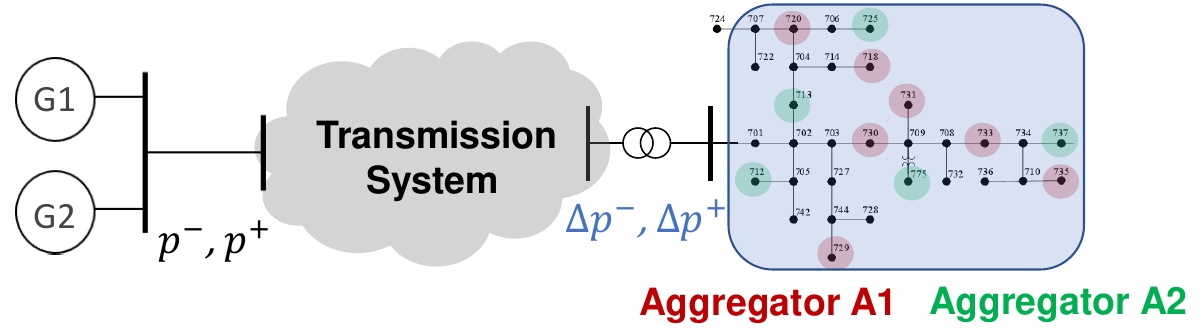}
		    \vspace{-.3cm}
	\caption{Example of transmission-distribution interaction.}
	\label{fig:DER_aggregate}
	    \vspace{-.5cm}
\end{figure}

Consider the grid depicted in Fig.~\ref{fig:DER_aggregate} where generators and a distribution grid are connected to the transmission system. On the transmission side of the grid, two generators G1 and G2 operated by separate companies bid into the electricity market. The amount of power provided by each of them is determined by the ISO by solving a transmission system dispatch problem. This problem considers the transmission line constraints defined by the minimum and maximum power transfer capacity~$p^-,p^+$, which bounds the total amount of power that the ISO can get from the two generators combined. This example is a simple illustration of the current state of the art in ISO market clearing.

Now, consider the distribution side of the grid where two DER aggregators A1 and A2 control DERs at multiple nodes, as illustrated by the red and green circles. These aggregators can now bid the flexibility offered by the DERs into the electricity market. However, including the operational limits of each individual DER (analogous to individual generator constraints) along with a detailed model of the distribution grid (analogous to the transmission grid model) into the optimization problem would significantly increase the computational complexity~\cite{li2016coordinated,lin2017decentralized,lin2016decentralized}. As an alternative, we propose a method to enable the
DSO to identify aggregate active power flexibility limits~$\Delta p^-,\Delta p^+$ around the current operating point. These limits represent the amount of active power flexibility that can be provided from the combined set of DERs (controlled both by A1 and A2) without causing internal distribution grid constraints to be violated. 
By integrating these limits in their dispatch algorithms, the ISO can determine a secure amount of power flexibility to procure from the aggregators.

\subsection{Bilevel Modeling Framework}
To enable the DSO to identify a secure aggregate power flexibility range, we propose a bilevel optimization approach.

\subsubsection{Bilevel problem overview}
The upper-level (or leader) problem represents the decision making of the DSO, and maximizes the aggregate change in DER power injections relative to the current operating point while guaranteeing that the worst-case voltage magnitudes (obtained by solving a set of lower level problems) stay within acceptable limits. The decision variables $\boldsymbol{X}_\mathbf{u}$ of this upper-level problem are the control actions determined by the DSO. Importantly, this set of decision variables include $\Delta \boldsymbol{p}^+$ and $\Delta \boldsymbol{p}^-$, the upper and lower limits on the aggregate active power flexibility, that can safely be provided from the feeder. 
Each lower-level (or follower) problem solves a three-phase optimal power flow (OPF) problem for a given node $i \in \mathcal{N}$ in the network with objective to find the worst-case (largest or smallest) voltage magnitude achievable at that node. Here, $\mathcal{N}$ represent the set of three-phase nodes in the feeder (excluding the substation, which acts as a slack node) and $n=3\cdot |\mathcal{N}|$ is the total number of single-phase nodes in the network. The lower-level problem formulation is similar to the model proposed in~\cite{molzahn2019grid}, which we extend to include the unbalanced three-phase power flow model derived in~\cite{girigoudar2021linearized}. The decision variables $\boldsymbol{X}_\mathbf{l}$ of the lower-level problem are assumed to be independently determined by the aggregator(s) and may therefore be equal to the values that would lead to worst case conditions in the grid.
Note that there are many lower level problems. Specifically, given a network with~$n$ single-phase nodes, we need~$2n$ lower-level problems, i.e. $n$ problems identifying smallest achievable voltage magnitude and~$n$ problems finding largest achievable voltage magnitude for each single-phase node.

\subsubsection{Input data requirements}
To solve the bilevel problem, we require information regarding the distribution grid parameters as well as information about the current operating state. Furthermore, the location and feasible range of set-points for individual DERs is assumed to be known. Note that this DER information does not have to be real-time information, but could simply be device characteristics provided by the aggregators when new DERs enroll in their programs.

\subsubsection{Rules on DER activation}
\label{sec:DER_rules}
In order to obtain practical ranges of flexibility which are not overly conservative, we enforce a simple rule on DER control actuation. This rule states that the active power injection of all DERs in the feeder have to either be increased or decreased (i.e., we cannot have a scenario where some DERs reduce their power injections and others increase them). This rule is needed to ensure that limiting the aggregate active power response is sufficient to limit voltage magnitude violations in the feeder. Without it, DERs in one part of the feeder can arbitrarily increase their injections as long as the increases are offset by an equally large decrease in injections elsewhere. If this is allowed, even a zero change in the overall active power injection at the substation can cause voltage violations. 
To integrate this rule in our optimization problem, we consider two cases:
\begin{enumerate}[(a)]
    \item \emph{Positive (+) case:}  The positive case assumes that the active power injections at any node~$i \in \mathcal{N}$ connected to phase~$\phi \in \{a,b,c\}$ is controlled to a value that is \textbf{greater} than the current active power injection. For loads, this implies that demand is reduced while for DERs such as solar PV inverters, it implies that active power generation is increased. Thus, we include the following constraints, 
     \begin{align}
         \Delta \boldsymbol{p}_{\mathbf{L},i}^\phi \leq 0, ~~
         \Delta \boldsymbol{p}_{\mathbf{G},i}^\phi \geq 0,~\forall_{\phi \in \{a,b,c\},i \in \mathcal{N}}. \label{eq:P_pos}
    \end{align}
    Here $\Delta \boldsymbol{p}_{\mathbf{L},i}^\phi, \Delta \boldsymbol{p}_{\mathbf{G},i}^\phi$ represent the deviation of the active power of the individual loads and solar PV inverters at node~$i\in \mathcal{N}$ connected to phase~$\phi \in \{a,b,c\}$, respectively, from the current operating point. 
    
    \item \emph{Negative (-) case:}  This negative case assumes that the active power injections at every node~$i \in \mathcal{N}$ connected to phase~$\phi$ is controlled to a value that is \textbf{lower} than the current operating point. This gives rise to the following constraints for flexibility activation,
     \begin{align}
         \Delta \boldsymbol{p}_{\mathbf{L},i}^\phi \geq 0, ~~
         \Delta \boldsymbol{p}_{\mathbf{G},i}^\phi \leq 0,~&\forall_{\phi \in \{a,b,c\},i \in \mathcal{N}}. \label{eq:P_neg}
    \end{align}
\end{enumerate}
Enforcing this rule increases the number of lower-level problems we need to solve, as the worst-case conditions determined for the positive and negative cases will be different. Given a network with~$n$ single-phase nodes, we now need~$4n$ lower-level problems, i.e. $2n$ each for the positive and negative case.

\subsubsection{DER reactive power control}
We further assume that all DERs are equipped with smart inverters which can provide reactive power support as per IEEE Standard 1547-2018~\cite{photovoltaics2018ieee} and that the DSO is allowed to determine the reactive power setpoints for these inverters. By making this assumption, it is possible to significantly increase the amount of aggregate power flexibility that DERs are allowed to provide, as demonstrated by our case study results in Section~\ref{sec:sim_res}. In this work, we consider three reactive power modes (i.e. constant power factor mode, constant reactive power mode and voltage-reactive power mode) in which the inverters can operate.
The detailed modelling of the inverter modes is discussed later in Section~\ref{subsec:inv_modes}.

    \section{Bilevel Problem Formulation}
\label{sec:prob_for}
In this section, we describe our model of the distribution grid and the flexibility offered by DERs such as solar PV inverters and controllable loads, before presenting the full bilevel problem formulation. 

\subsection{Notation}
We consider a network where the distribution substation with index~$i=0$ is chosen as the slack node.~$\mathcal{N}$ denotes the set of non-slack nodes. Without loss of generality, we assume all nodes have three phases with the set of phases defined by~$\Phi = \{a,b,c\}$, and total number of single-phase nodes is~$n=3 \cdot |\mathcal{N}|$.  Apart from the substation, the only other generators in the network are single-phase, residential solar PV systems. To simplify notation, we assume that there is one solar PV inverter and one load at each single-phase node. If there is no solar PV or load at some node, the corresponding entries are set to zero. All the optimization variables are denoted using bold symbols and all vectors as well as matrices are represented using capital letters. Given a vector~$X \in \mathbb{C}^n$, the matrix ~${dg}(X)\in \mathbb{C}^{(n \times n)}$~is a diagonal matrix with elements of~$X$~on its diagonal. We use~$\odot$ to denote the element-wise product of two vectors. Given a complex phasor~$x$, the complex conjugate is represented by~$x^*$. 


\subsection{Current Operating Point}
At any three-phase node~$i \in \mathcal{N}$, we assume that we have access to the current voltage phasor~${V}_i=V_{\text{d}i}+j V_{\text{q}i}$, with~$V_{\text{d}i}=[v_{\text{d}i}^a ~~v_{\text{d}i}^b ~~v_{\text{d}i}^c]^\top$ and~$V_{\text{q}i}=[v_{\text{q}i}^a ~~v_{\text{q}i}^b ~~v_{\text{q}i}^c]^\top$ representing the real and imaginary components, respectively. Similarly, we can define the three-phase active and reactive power injections by~${P}_i=[p_i^a ~~p_i^b ~~p_i^c]^\top$ and~${Q}_i=[q_i^a ~~q_i^b ~~q_i^c]^\top$, respectively. The power injections are expressed as the difference between the generation and load demand using
\begin{subequations}
\begin{align}
     {p}_{i}^\phi = {p}_{\text{G},i}^\phi-{p}_{\text{L},i}^\phi,~&\forall_{\phi \in \Phi,i \in \mathcal{N}}, \\
     {q}_{i}^\phi = {q}_{\text{G},i}^\phi-{q}_{\text{L},i}^\phi,~&\forall_{\phi \in \Phi,i \in \mathcal{N}}, \label{eq:PQ_inj} 
\end{align}
\end{subequations}
where~${p}_{\text{G},i}^\phi,{q}_{\text{G},i}^\phi$ are the active and reactive power generation by solar PV inverters;~${p}_{\text{L},i}^\phi,{q}_{\text{L},i}^\phi$ denote the respective active and reactive load demand. 
The power balance is maintained by the substation, which supplies the difference between the load and power generation as well as other losses in the grid. We denote the current active and reactive power injection at the substation by~${P}_{\text{G},0},{Q}_{\text{G},0}\in \mathbb{R}^{3}$. Note that the assumption of perfect knowledge of the current system state can be relaxed to include a range of possible operating points. 

\subsection{System Modeling in Lower-Level Problem}
We next describe the model used in the lower-level problem to represent the worst-case operating condition after DER control actuation.  

\subsubsection{Voltage representation}
The voltage variables in the bilevel problem are expressed in rectangular form. For any node~$i\in \{0,\mathcal{N}\}$, the real and imaginary components of voltage are given by~$\boldsymbol{V}_{\mathbf{d}i}=[\boldsymbol{v}_{\mathbf{d}i}^a ~~\boldsymbol{v}_{\mathbf{d}i}^b ~~\boldsymbol{v}_{\mathbf{d}i}^c]^\top$ and~$\boldsymbol{V}_{\mathbf{q}i}= [\boldsymbol{v}_{\mathbf{q}i}^a ~~\boldsymbol{v}_{\mathbf{q}i}^b ~~\boldsymbol{v}_{\mathbf{q}i}^c]^\top$, respectively. The distribution substation is assumed to be perfectly balanced, i.e.
\begin{align}
\boldsymbol{V}_{\mathbf{d}0} +j \cdot \boldsymbol{V}_{\mathbf{q}0} &= 
    \begin{bmatrix}
        e^{j0^\circ} & e^{-j120^\circ} &  e^{j120^\circ}
    \end{bmatrix}^\top. \label{eq:Vref}
\end{align}
For every other node~$i \in \mathcal{N}$ with phase~$\phi$, the voltage magnitude variable is given by~
$|\boldsymbol{v}_{i}^\phi|=\sqrt{(\boldsymbol{v}_{\mathbf{d}i}^\phi)^2 + (\boldsymbol{v}_{\mathbf{q}i}^\phi)^2}$.
To avoid introducing a non-linear constraint for the voltage magnitude, we introduce a linear approximation for this relationship. Given the current voltage magnitude~$|{v}_i^\phi| = \sqrt{{(v_{\text{d}i}^\phi)}^2+{(v_{\text{q}i}^\phi)}^2}$, we use a First-order Taylor approximation to get the following linear equation:
\begin{align}
\label{eq:Vlim_lin}
  &{(v_{\text{d}i}^\phi)}^2 \!+\! {(v_{\text{q}i}^\phi)}^2 \!+\! 2v_{\text{d}i}^\phi\boldsymbol{v}_{\mathbf{d}i}^\phi \!+\! 2v_{\text{q}i}^\phi\boldsymbol{v}_{\mathbf{q}i}^\phi = {{|{v}_i^\phi|}^2 \!+ \! 2|{v}_i^\phi||\boldsymbol{v}_i^\phi|}.
\end{align}
Note that we do not enforce voltage magnitude constraints in the lower level problem, as these are accounted for in the upper level problem, as further discussed in Section \ref{sec:bilevel_prob}.


\subsubsection{Load modeling}
The total demand from load  at node~$i$ in phase $\phi$ is constrained by the lower and upper limits ~$\underline{p}_{\text{L},i}^\phi,\overline{p}_{\text{L},i}^\phi$ for the active power demand from this node, i.e.
\begin{align}
     \underline{p}_{\text{L},i}^\phi \leq {p}_{\text{L},i}^\phi + \Delta \boldsymbol{p}_{\mathbf{L},i}^\phi \leq \overline{p}_{\text{L},i}^\phi,~\forall_{\phi \in \Phi,i \in \mathcal{N}}.  \label{eq:P_L} 
\end{align}
Assuming that loads operate with a constant power factor~$pf_{\text{L},i}^{\phi}$, the reactive power demand is given by
\begin{align}
    & \boldsymbol{q}_{\mathbf{L},i}^\phi ={\frac{\sqrt{1-{(pf_{\text{L},i}^\phi)}^2}}{{pf_{\text{L},i}^\phi}}}  \cdot \left({p}_{\text{L},i}^\phi + \Delta \boldsymbol{p}_{\mathbf{L},i}^\phi \right),~\forall_{\phi \in \Phi,i \in \mathcal{N}}. \label{eq:load_q}
\end{align}

\subsubsection{Solar PV active power modeling}
The constraint on solar PV generation at node $i$ in phase $\phi$ is given by
\begin{align}
     \underline{p}_{\text{G},i}^\phi \leq {p}_{\text{G},i}^\phi + \Delta \boldsymbol{p}_{\mathbf{G},i}^\phi \leq \overline{p}_{\text{G},i}^\phi,~\forall_{\phi \in \Phi,i \in \mathcal{N}},  \label{eq:P_G} 
\end{align}
where~$\underline{p}_{\text{G},i}^\phi,\overline{p}_{\text{G},i}^\phi$ denote the respective lower and upper limits for the active power generation of the inverter. The reactive power modeling of solar PV inverters is discussed later. 

\subsubsection{Aggregate active power flexibility}
The total active power flexibility~$\Delta \boldsymbol{p}$ provided by the distribution grid is given by the sum of the flexibility from individual loads and solar PV inverters, and is bounded by lower and upper limits of the aggregate power flexibility provided from the feeder $\Delta \boldsymbol{p}^-,\Delta \boldsymbol{p}^+$. This gives rise to the following constraint,  
\begin{align}
     &\Delta \underline{p} \!\leq\! \Delta \boldsymbol{p}^- \!\leq\! \underbrace{\sum_{i \in \mathcal{N}}\sum_{\phi \in \Phi}\left(\Delta \boldsymbol{p}_{\mathbf{G},i}^\phi - \Delta \boldsymbol{p}_{\mathbf{L},i}^\phi \right)}_{\Delta \boldsymbol{p}} \!\leq \! \Delta \boldsymbol{p}^+ \! \leq \! \Delta \overline{p}.  \label{eq:P_ss_constr} 
\end{align}
Here, $\Delta \overline{p}={\sum_{i \in \mathcal{N}}\sum_{\phi \in \Phi}(\overline {p}_{\text{G},i}^\phi - {p}_{\text{G},i}^\phi + \overline {p}_{\text{L},i}^\phi- {p}_{\text{L},i}^\phi )}$ is the maximum available aggregate flexibility that can be offered by the network, and $\Delta \underline{p}={\sum_{i \in \mathcal{N}}\sum_{\phi \in \Phi}(\underline {p}_{\text{G},i}^\phi - {p}_{\text{G},i}^\phi + \underline {p}_{\text{L},i}^\phi- {p}_{\text{L},i}^\phi )}$ is the minimum available aggregate flexibility.
In each lower level problem, the constraint \eqref{eq:P_ss_constr} is combined with the condition that the active power injections either increase \eqref{eq:P_pos} or decrease \eqref{eq:P_neg}. Note that setting~$\Delta \boldsymbol{p}=0$ forces $\Delta \boldsymbol{p}_{\mathbf{L},i}^\phi=\Delta \boldsymbol{p}_{\mathbf{G},i}^\phi=0$ and recovers the current operating point. 

\subsubsection{Power Flow}
We use a fixed-point power flow interpretation in this paper to linearize the power flow equations because it exhibits better global approximation accuracy when compared to other linear models such as first-order Taylor approximation~\cite{bernstein2018load}. With this approximation, the vector of voltage variables~$\boldsymbol{V}_\mathbf{d} = [ \boldsymbol{V}_{\mathbf{d}i \in \mathcal{N}}^\top]^\top,\boldsymbol{V}_\mathbf{q} = [ \boldsymbol{V}_{\mathbf{q}i \in \mathcal{N}}^\top]^\top$ is expressed as
\begin{subequations}
\label{eq:FP_rec}
\begin{align}
    &\boldsymbol{V}_\mathbf{d}  = \Re\Big\{{Z}_1 \Big\} + \Re\Big\{Z_2 \Big\} \cdot \boldsymbol{P} + \Im\Big\{Z_2 \Big\} \cdot \boldsymbol{Q}, \\
    &\boldsymbol{V}_\mathbf{q}   = \Im\Big\{{Z}_1 \Big\} + \Im\Big\{Z_2 \Big\} \cdot \boldsymbol{P} - \Re\Big\{Z_2 \Big\} \cdot \boldsymbol{Q}.
\end{align} 
\end{subequations}
Here,~$Z_1 \in \mathbb{R}^n$~and~$Z_2 \in \mathbb{R}^{n\times n}$~are fixed and calculated using the current operating point as described in~\cite{girigoudar2021linearized}.~$\boldsymbol{P} \in \mathbb{R}^n$~includes the active power flexibility~$\Delta \boldsymbol{P}_\mathbf{L} = [ \Delta \boldsymbol{P}_{\mathbf{L}i \in \mathcal{N}}^\top]^\top,\Delta \boldsymbol{P}_\mathbf{G} = [ \Delta \boldsymbol{P}_{\mathbf{G}i \in \mathcal{N}}^\top]^\top$ from loads and PV inverters, respectively, and~$\boldsymbol{Q} \in \mathbb{R}^n$ includes the reactive power injections of PV inverters~$ \boldsymbol{Q}_\mathbf{G} = [  \boldsymbol{Q}_{\mathbf{G}i \in \mathcal{N}}^\top]^\top$ and load reactive power demand~$ \boldsymbol{Q}_\mathbf{L} = [  \boldsymbol{Q}_{\mathbf{L}i \in \mathcal{N}}^\top]^\top$, as defined by
\begin{subequations}
\begin{align}
    &\boldsymbol{P}= {P}_\text{G}+ \Delta \boldsymbol{P}_\mathbf{G} - ({P}_\text{L}+ \Delta \boldsymbol{P}_\mathbf{L}), \\
    &\boldsymbol{Q}= \boldsymbol{Q}_\mathbf{G} -  \boldsymbol{Q}_\mathbf{L},
\end{align}    
\end{subequations}
where~$P_\text{G}=[P_{\text{G}i \in \mathcal{N}}^\top]^\top,P_\text{L}=[P_{\text{L}i \in \mathcal{N}}^\top]^\top$ are the respective active power generation and load demand vectors at the current operating point.

\subsection{Reactive Power Control from Solar PV Inverters}
\label{subsec:inv_modes}
Following the IEEE Standard 1546-2018~\cite{photovoltaics2018ieee}, we consider three modes in which the PV inverters can operate. 
We next describe some constraints that are common to all inverters, before discussing the three reactive power control modes. Note that the set-points provided by the DSO are different for each reactive power control mode.

\subsubsection{Inverter Constraints}
For all three modes, the solar PV inverter at node~$i$ connected to phase~$\phi$ should not exceed the inverter apparent power capacity~$|{s_{\text{G},i}^\phi}|$. In its original form, the apparent power constraint is quadratic. To avoid introducing non-linear constraints, we outer approximate the circular feasible region using linear constraints~\cite{chen2015robust}. The resulting constraints are given by 
\begin{subequations}
\label{eq:inv_lim_lin}
\begin{align}
& 0 \leq p_{\text{G},i}^\phi+ \Delta \boldsymbol{p}_{\mathbf{G},i}^\phi   \leq |{s_{\text{G},i}^\phi}|,&\forall_{\phi \in \Phi,i \in \mathcal{N}}, \\
   &-|{s_{\text{G},i}^\phi}| \leq  \boldsymbol{q}_{\mathbf{G},i}^\phi   \leq |{s_{\text{G},i}^\phi}|,&\forall_{\phi \in \Phi,i \in \mathcal{N}}, \\
   & p_{\text{G},i}^\phi+ \Delta \boldsymbol{p}_{\mathbf{G},i}^\phi + \boldsymbol{q}_{\mathbf{G},i}^\phi   \leq \sqrt{2} \cdot|{s_{\text{G},i}^\phi}|,&\forall_{\phi \in \Phi,i \in \mathcal{N}}, \\ 
      & p_{\text{G},i}^\phi+ \Delta \boldsymbol{p}_{\mathbf{G},i}^\phi - \boldsymbol{q}_{\mathbf{G},i}^\phi   \leq \sqrt{2} \cdot |{s_{\text{G},i}^\phi}|,&\forall_{\phi \in \Phi,i \in \mathcal{N}}.
\end{align}
\end{subequations}

\subsubsection{Constant power factor mode}
This mode assumes that the DSO provides a target power factor setting~$pf_{\text{G},i}^\phi$ to the DERs. The reactive power injection~$\boldsymbol{q}_{\mathbf{G},i}^\phi$ is given by
\begin{subequations}
\label{eq:inv_cons_pf}
\begin{align}
    &\boldsymbol{q}_{\mathbf{G},i}^\phi  =\boldsymbol{\gamma}_{\mathbf{G},i}^\phi \cdot \left({p}_{\text{G},i}^\phi + \Delta \boldsymbol{p}_{\mathbf{G},i}^\phi \right), \hspace{4.7em} \forall_{\phi \in \Phi,i \in \mathcal{N}},\\
     & \!{\frac{-\sqrt{1 \!-{(pf_{\text{G},i}^\phi)}^2}}{{pf_{\text{G},i}^\phi}}} \!\leq\! \boldsymbol{\gamma}_{\mathbf{G},i}^\phi \!\leq\!  {\frac{\sqrt{1 \!-{(pf_{\text{G},i}^\phi)}^2}}{{pf_{\text{G},i}^\phi}}},~\forall_{\phi \in \Phi,i \in \mathcal{N}},
    \end{align}
\end{subequations}
where~$\boldsymbol{\gamma}_{\mathbf{G},i}^\phi$ is the power ratio of the inverter and is a decision variable in the upper-level optimization problem. Note that~(\ref{eq:inv_cons_pf}a) consist of bilinear terms involving the upper-level variables~$\boldsymbol{\gamma}_{\mathbf{G},i}^\phi$ and lower-level variables~$\Delta \boldsymbol{p}_{\mathbf{G},i}^\phi$. For a bilevel problem, the upper-level variables are considered as parameters in the lower-level problem, and as a result,~\eqref{eq:inv_cons_pf} is linear for the lower-level problem. 

\subsubsection{Constant reactive power mode}
In this mode, the reactive power injection~$\boldsymbol{q}_{\mathbf{G},i}^\phi$ is specified by the DSO and hence, considered as an upper-level variable. In our work, we assume that PV inverters are operating within some power factor range~\cite{photovoltaics2018ieee} defined by the power ratio~$\gamma_{\text{G},i}^\phi$ and we enforce limits on the reactive power using
\begin{subequations}
\label{eq:inv_cons_q}
\begin{align}
  & \boldsymbol{q}_{\mathbf{G},i}^\phi \geq -\gamma_{\text{G},i}^\phi \! \cdot \! \left({p}_{\text{G},i}^\phi + \Delta \boldsymbol{p}_{\mathbf{G},i}^\phi \right),&\forall_{\phi \in \Phi,i \in \mathcal{N}}, \\
  &\boldsymbol{q}_{\mathbf{G},i}^\phi \leq \gamma_{\text{G},i}^\phi \! \cdot \! \left({p}_{\text{G},i}^\phi + \Delta \boldsymbol{p}_{\mathbf{G},i}^\phi \right),&\forall_{\phi \in \Phi,i \in \mathcal{N}}. 
\end{align}
\end{subequations}

\subsubsection{Voltage-reactive power mode}
\begin{figure}[!h]
	\centering	  
		    \vspace{-.3cm}
	\includegraphics[width=0.24\textwidth]{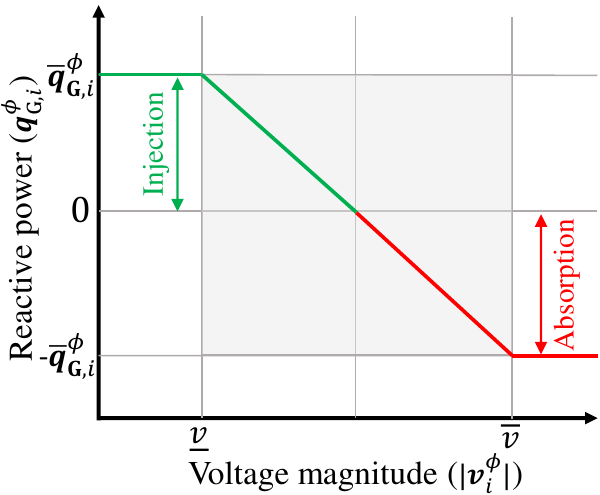}
	    \vspace{-.3cm}
	\caption{Reactive power-voltage magnitude characteristic.}
	\label{fig:QVcurve}
	    \vspace{-.5cm}
\end{figure}

In this mode, the reactive power is a function of the voltage magnitude, as illustrated in Fig.~\ref{fig:QVcurve} for a PV inverter at node~$i$ connected to phase~$\phi$. The maximum available reactive power~$\overline{\boldsymbol{q}}_{\mathbf{G},i}^\phi$ is specified by the DSO and hence, included in our bilevel task as an upper-level decision variable. When operating outside the grey shaded region in Fig.~\ref{fig:QVcurve}, where the voltage magnitude is violating the specified limits~$\underline{v},\overline{v}$, the inverter either injects or absorbs the maximum available reactive power~$\overline{\boldsymbol{q}}_{\mathbf{G},i}^\phi$. In the grey shaded region, the reactive power~$\boldsymbol{q}_{\mathbf{G},i}^\phi$ depends on the voltage magnitude~$|\boldsymbol{v}_i^\phi|$ and can be expressed as
\begin{subequations}
\label{eq:inv_qv}
\begin{align}
    & \boldsymbol{q}_{\mathbf{G},i}^\phi = \overline{\boldsymbol{q}}_{\mathbf{G},i}^\phi-2 \overline{\boldsymbol{q}}_{\mathbf{G},i}^\phi \cdot \left( \frac{|\boldsymbol{v}_i^\phi|-\underline{v}}{\overline{v}-\underline{v}}\right),&\forall_{\phi \in \Phi,i \in \mathcal{N}},\\
    &0 \leq \overline{\boldsymbol{q}}_{\mathbf{G},i}^\phi \leq |s_{\text{G},i}^\phi|, &\forall_{\phi \in \Phi,i \in \mathcal{N}}.
\end{align}
\end{subequations}
Here, the actual upper bound of the maximum available reactive power~$\overline{\boldsymbol{q}}_{\mathbf{G},i}^\phi$ is given by~$\sqrt{|s_{\text{G},i}^\phi|^2 - ({p}_{\text{G},i}^\phi + \Delta \boldsymbol{p}_{\mathbf{G},i}^\phi)^2}$. Since this expression is nonlinear, we relax the upper bound as defined in~(\ref{eq:inv_qv}b). 

We assume that the inverter is always operating in the gray shaded region shown in Fig.~\ref{fig:QVcurve}. This assumption is reasonable since we enforce constraints on the voltage magnitude~$|\boldsymbol{v}_i^\phi|$ to be within the limits~$\underline{v},\overline{v}$ in the upper-level problem. Similar to the constant power factor mode, the reactive power constraints in~(\ref{eq:inv_qv}a) consist of bilinear terms involving the upper-level variables~$\overline{\boldsymbol{q}}_{\mathbf{G},i}^\phi$ and lower-level variables~$|\boldsymbol{v}_i^\phi|$.

\subsection{Bilevel Optimization Problem}
\label{sec:bilevel_prob}
The objective of the bilevel problem is to find the maximum range of aggregate power flexibility while ensuring that the grid is secure even in the worst-case conditions. In this section, we discuss the procedure to formulate the bilevel problem when inverters are operating in constant power factor mode and DSO controls the power ratio~$\boldsymbol{\gamma}_{\mathbf{G},i}^\phi$ of each inverter. The same procedure can be followed for bilevel problem formulations with inverters operating in other reactive power modes. The  upper-level variable vector~$\boldsymbol{X}_\mathbf{u}$ and lower-level variable vector~$\boldsymbol{X}_\mathbf{l}$ are denoted by
\begin{align}
    & \boldsymbol{X}_\mathbf{u} := \{\Delta \boldsymbol{p}^+,\Delta \boldsymbol{p}^-,\boldsymbol{\gamma}_{\mathbf{G},i}^\phi\forall_{\phi \in \Phi,i \in \mathcal{N}} \}, \notag \\
    &\boldsymbol{X}_\mathbf{l} := \left\{\{\boldsymbol{v}_{\mathbf{d}i}^\phi,  \boldsymbol{v}_{\mathbf{q}i}^\phi, |{\boldsymbol{v}_i^\phi}|,  \Delta \boldsymbol{p}_{\mathbf{G},i}^\phi,  \Delta \boldsymbol{p}_{\mathbf{L},i}^\phi,  \boldsymbol{q}_{\mathbf{L},i}^\phi,\boldsymbol{q}_{\mathbf{G},i}^\phi \}_{\forall_{\phi \in \Phi,i \in \mathcal{N}}}\right\}. \notag
\end{align}
Since we need to represent both the positive and negative active power activation cases and must also determine worst-case voltage magnitude (minimum and maximum) achievable, we need to consider 4 lower level problems at every single-phase node. Therefore, we use four sets of lower-level variables corresponding to  four scenarios summarized in Table~\ref{tab:var_ll}. 

Each of the lower-level problems include system, inverter and active power flexibility constraints. The set of system constraints is given by
\begin{align}
    \mathbb{S}({\boldsymbol{X}}_\mathbf{l})
    :=
    \begin{cases}
        \text{Voltage constraints}~\eqref{eq:Vlim_lin}, \\
        \text{Load reactive power constraints}~\eqref{eq:load_q}, \\
        \text{Power flow}~\eqref{eq:FP_rec}.
    \end{cases} \notag
\end{align}
The set of inverter constraints (for the constant power factor mode) is given by
\begin{align}
    \mathbb{I}(\boldsymbol{\gamma}_{\mathbf{G},i}^\phi,{\boldsymbol{X}}_\mathbf{l})
    := \text{Inverter constraints}~\eqref{eq:inv_lim_lin},\eqref{eq:inv_cons_pf}, \notag
\end{align}
while the active power flexibility constraints for either positive or negative case are represented by
\begin{align}
    &\mathbb{F}(\Delta \boldsymbol{p}^+,\Delta \boldsymbol{p}^-,{\boldsymbol{X}}_\mathbf{l})
 :=  \text{Positive case}~\eqref{eq:P_pos},\eqref{eq:P_ss_constr}~\text{or} \notag \\
     &\mathbb{F}(\Delta \boldsymbol{p}^+,\Delta \boldsymbol{p}^-,{\boldsymbol{X}}_\mathbf{l})
 :=  \text{Negative case}~\eqref{eq:P_neg},\eqref{eq:P_ss_constr}. \notag
\end{align}

\begin{table}[!t]
\centering
\caption{Notation of lower-level problem variables corresponding to different scenarios}
  \vspace{-.2cm}
\begin{tabular}{cccc}
\toprule
\multirow{1}{*}{\textbf{Scenario}}  & \multirow{1}{*}{\textbf{Active power}} &  \multirow{1}{*}{\textbf{Worst case}} &  \multirow{1}{*}{\textbf{Lower-level}}   \\
\textbf{(\#)} & \textbf{activation case} & \textbf{voltage magnitude} & \textbf{variable set}\\
\midrule \midrule
1 & {Positive} & Minimum  & ${\boldsymbol{X}}_{\mathbf{l},1}$ \\ \midrule
2 & Positive & Maximum  & ${\boldsymbol{X}}_{\mathbf{l},2}$ \\ \midrule
3 & {Negative} & Minimum  & ${\boldsymbol{X}}_{\mathbf{l},3}$ \\ \midrule
4 & Negative & Maximum  & ${\boldsymbol{X}}_{\mathbf{l},4}$ \\ 
\bottomrule
\end{tabular}
  \vspace{-.5cm}
\label{tab:var_ll}
\end{table}

The bilevel problem to determine the aggregate power flexibility limits~$\Delta \boldsymbol{p}^+,\Delta \boldsymbol{p}^-$ can then be formulated as
\begin{align}
   \max_{\boldsymbol{X}_\mathbf{u}}~ & \Delta{\boldsymbol{p}^+} - \Delta{\boldsymbol{p}^-}   \tag{$\text{P}^\pm$}\\
    \text{s.t. } &  \underline{v} \leq |\boldsymbol{{v}}_{i}^\phi|_1, |\boldsymbol{{v}}_{i}^\phi|_2\leq \overline{v}, \hspace{4.5em} \forall_{\phi \in \Phi, i \in \mathcal{N}},  \notag\\
    &  \underline{v} \leq |\boldsymbol{{v}}_{i}^\phi|_3, |\boldsymbol{{v}}_{i}^\phi|_4\leq \overline{v}, \hspace{4.5em} \forall_{\phi \in \Phi, i \in \mathcal{N}},  \notag\\
     & \hspace{-0.8cm}\text{where}~\forall {s \in \{1,2,3,4\}}, \notag \\
     & \hspace{-0.8cm} |\boldsymbol{{v}}_{i}^\phi|_s=\max_{{\boldsymbol{X}}_{\mathbf{l},s}}~ (-1)^s \cdot |\boldsymbol{{{v}}}_{i}^\phi|, \hspace{4.5em} \forall_{\phi \in \Phi, i \in \mathcal{N}}, \notag \\
     & \hspace{1.9em}  \text{s.t. } \, \mathbb{S}({\boldsymbol{X}}_{\mathbf{l},s}), ~    \mathbb{I}(\boldsymbol{\gamma}_{\mathbf{G},i}^\phi,{\boldsymbol{X}}_{\mathbf{l},s}), \notag \\
    & \hspace{3.6em}    \mathbb{F}(\Delta \boldsymbol{p}^+,\Delta \boldsymbol{p}^-,{\boldsymbol{X}}_{\mathbf{l},s}), \notag 
\end{align}
Similar problems can be formulated for other types of inverter control by replacing $\mathbb{I}(\boldsymbol{\gamma}_{\mathbf{G},i}^\phi,{\boldsymbol{X}}_\mathbf{l})$ with either
\eqref{eq:inv_lim_lin},~\eqref{eq:inv_cons_q} for constant reactive power mode or \eqref{eq:inv_lim_lin},~\eqref{eq:inv_qv} for voltage-reactive power mode.

    \section{Problem Reformulation and Solution Method}
\label{sec:prob_refor}
We next use the strong-duality theorem for the lower-level problem to reduce the bilevel problem into a single-level problem. Given that the lower-level problem has a finite optimal solution, the dual feasible set is non empty and strong duality holds for every primal and dual feasible pairs. 

\subsection{Single-level Reformulation}
To derive the strong duality based reformulation of the bilevel problem~$\text{P}^\pm$, we first rewrite the bilevel problem in a simplified and more abstract form with one follower to get
\begin{subequations}
\label{eq:gen_bilevel}
\begin{align}
   \max_{\Delta{\boldsymbol{p}},\boldsymbol{\gamma}}~ & \Delta{\boldsymbol{p}}    \\
    \text{s.t. } &  \boldsymbol{\hat{x}}_\mathbf{l} \leq b_1,  \\
& \text{where}~\boldsymbol{\hat{x}}_\mathbf{l}=\max_{\boldsymbol{x}_\mathbf{l}}~  \boldsymbol{x}_\mathbf{l}    \\
     & \hspace{5.9em} \text{s.t. }  a_2 \boldsymbol{x}_\mathbf{l} \leq b_2,  \\
     & \hspace{7.5em} a_3\boldsymbol{x}_\mathbf{l} + d_3 \boldsymbol{\gamma} \boldsymbol{x}_\mathbf{l} \leq b_3,  \\
& \hspace{7.5em}  a_4 \boldsymbol{x}_\mathbf{l} +d_4 \Delta{\boldsymbol{p}} \leq b_4,  
\end{align}
\end{subequations}
where~$\boldsymbol{x}_\mathbf{l}$ is the follower problem variable and~$\boldsymbol{\gamma},\Delta{\boldsymbol{p}}$ represent the upper-level variables. Therefore, the follower   problem~(\ref{eq:gen_bilevel}c)-(\ref{eq:gen_bilevel}f) is linear in~$\boldsymbol{x}_\mathbf{l}$ even though we have bilinear terms in~(\ref{eq:gen_bilevel}e).

By taking dual of the lower-level problem~(\ref{eq:gen_bilevel}c)-(\ref{eq:gen_bilevel}f) and applying the strong duality theorem~\cite{zeng2014solving}, we reduce the bilevel problem~\eqref{eq:gen_bilevel} to  a single-level optimization problem given by
\begin{subequations}
\label{eq:gen_bilevel_reform}
\begin{align}
   \max_{\Delta{\boldsymbol{p}},\boldsymbol{\gamma},\boldsymbol{x}_\mathbf{l},\boldsymbol{\lambda}}~ & \Delta{\boldsymbol{p}}    \\
     \text{s.t. } &  \boldsymbol{{x}}_\mathbf{l} \leq b_1,  \\
      & a_2 \boldsymbol{x}_\mathbf{l} \leq b_2,  \\
     &  a_3\boldsymbol{x}_\mathbf{l} + d_3 \boldsymbol{\gamma} \boldsymbol{x}_\mathbf{l} \leq b_3,  \\
&   a_4 \boldsymbol{x}_\mathbf{l} +d_4 \Delta{\boldsymbol{p}} \leq b_4,  \\
 & \boldsymbol{x}_\mathbf{l} \geq {\begin{bmatrix}
        b_2 & b_3 & (b_4-d_4\Delta\boldsymbol{p})
    \end{bmatrix}} \boldsymbol{\lambda},    \\
    & {\begin{bmatrix}
        a_2 & (a_3+d_3\boldsymbol{\gamma}) & a_4
    \end{bmatrix}} \boldsymbol{\lambda} = 1,  \\
    & \boldsymbol{\lambda} \geq 0,
\end{align}
\end{subequations}
where~(\ref{eq:gen_bilevel_reform}c)-(\ref{eq:gen_bilevel_reform}e) and~(\ref{eq:gen_bilevel_reform}g)-(\ref{eq:gen_bilevel_reform}h) are the primal and dual constraints for the lower level problem, respectively, and~(\ref{eq:gen_bilevel_reform}f) enforces strong duality.

A similar approach can be used to derive the single-level reduction of~$\text{P}^\pm$ which comprises of multiple follower problems. We can also choose other inverter reactive power control modes and formulate the bilevel problem and the single-level reformulated problem in the same way as described above for the constant power factor mode. Note that the presence of bilinear terms~$\boldsymbol{\gamma}\boldsymbol{x}_\mathbf{l}$ in the primal follower problem constraint~(\ref{eq:gen_bilevel_reform}d), $\Delta\boldsymbol{p} \boldsymbol{\lambda}$ in the strong duality constraint~(\ref{eq:gen_bilevel_reform}f) and~$\boldsymbol{\gamma}\boldsymbol{\lambda}$ in the dual equality constraint~(\ref{eq:gen_bilevel_reform}g) make the problem challenging to solve. We will next discuss an iterative solution approach to handle the bilinear terms and solve~\eqref{eq:gen_bilevel_reform} efficiently.

\subsection{Solution Method}
\label{sec:sol_approach}

As the number of follower problems in~\eqref{eq:gen_bilevel} increases, the number of bilinear (and non-convex) terms seen in the single-level reformulated problem~\eqref{eq:gen_bilevel_reform} also increase. Typically, the voltage magnitude violations in a distribution grid occur only at a few nodes, which in turn determine the aggregate power flexibility of the system. Thus, instead of solving the bilevel problem with all~$4n$ follower problems included, we 
use an iterative approach where we identify the nodes where voltage violations are most likely to occur and only include the lower-level problems corresponding to these nodes in the bilevel optimization problem~\eqref{eq:gen_bilevel_reform}.
The iterative approach is illustrated in Fig.~\ref{fig:sol_method} and comprises of the following steps:

\begin{figure}[!t]
	\centering	    
		  \vspace{-.2cm}
	\includegraphics[width=0.45\textwidth]{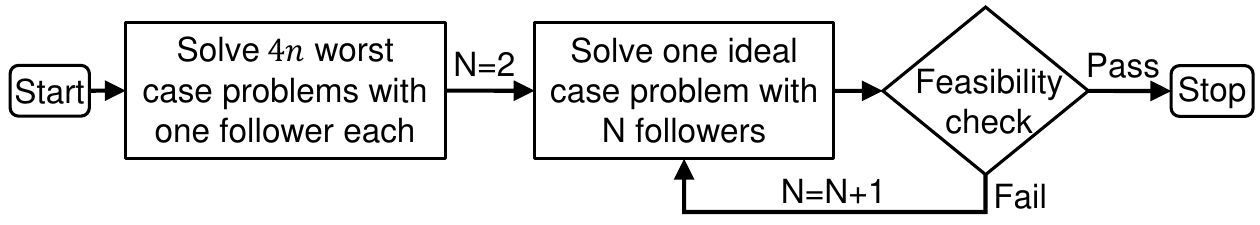}
	    \vspace{-.3cm}
	\caption{Iterative solution method to solve bilevel problem.}
	\label{fig:sol_method}
	   \vspace{-.5cm}
\end{figure}

\setcounter{figure}{4}
\begin{figure*}[!b]
	\centering	  
		  \vspace{-.2cm}
	\includegraphics[width=0.80\textwidth]{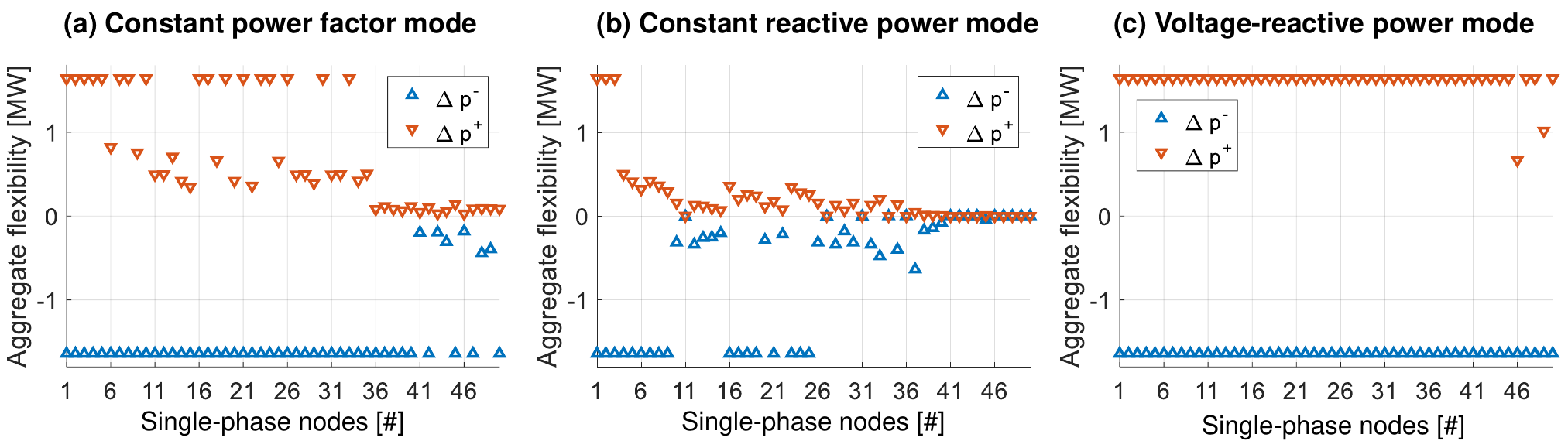}
	    \vspace{-.3cm}
	\caption{Worst-case upper (orange) and lower (blue) aggregate power flexibility limits for every single-phase node in the IEEE-13 node feeder. From left to right: Results of different inverter control modes.}
	\label{fig:worst_case}
	    \vspace{-.5cm}
\end{figure*}

    
    

\subsubsection{Worst case}
We first identify the aggregate flexibility range by assuming that the inverter set-points can take on worst-case values (instead of being chosen by the DSO). In this case, the inverter set-points become lower level variables.  
Solving the resulting problem is straightforward for the constant reactive power mode with linear inverter constraints~\eqref{eq:inv_cons_q}. However, the inverter constraints~(\ref{eq:inv_cons_pf}a),~(\ref{eq:inv_qv}a) for constant power factor and voltage-reactive power modes include bilinear terms for this case. To avoid this problem of non-convexity, we fix the inverter set-points. If the lower-level objective is to maximize voltage magnitude, the worst-case inverter setpoints are determined as:
\begin{itemize}
\item \emph{Constant power factor mode}: We set~$\boldsymbol{\gamma}_{\mathbf{G},i}^\phi$ for every inverter to its upper limit defined in~(\ref{eq:inv_cons_pf}b). This corresponds to the situation where all inverters are boosting the voltage profile in the network by injecting the full available reactive power, leading to higher voltage magnitudes.
\item \emph{Voltage-reactive power mode}: We set~$\overline{\boldsymbol{q}}_{\mathbf{G},i}^\phi$ to its lower limit defined in~(\ref{eq:inv_qv}b). This represents the condition where inverters provide no reactive power support when voltage magnitude is close to or violating the upper limit~$\overline{v}$.
\end{itemize}
If the lower-level objective is to minimize the voltage magnitude, we set the corresponding variables to the other extreme. 

With these assumptions,~$\Delta \boldsymbol{p}^+,\Delta \boldsymbol{p}^-$ are the only variables that are shared by all the lower-level problems, i.e. there are no other variables or constraints linking the~$4n$ follower problems. This allows us to solve each follower problem separately. Specifically, we solve~$4n$ worst case problems with one leader and one follower each to obtain~$2n$ aggregate power flexibility upper limits for the positive case (with $\Delta \boldsymbol{p}^-=0$) and~$2n$ aggregate power flexibility lower limits for the negative case (with $\Delta \boldsymbol{p}^+=0$). The difference between the minimum upper limit and the maximum lower limit is the worst case aggregate power flexibility range for the network corresponding to the worst-case inverter control. Note that each of the~$4n$ small worst case problems has only one bilinear term in the strong duality inequality constraint. It is faster to solve the~$4n$ smaller problems than one large problem with~$4n$ followers, especially if solved in parallel. 

\subsubsection{Ideal case}
Next, we solve the ideal case problem~P$^\pm$ as defined in Section~\ref{sec:bilevel_prob}. This problem cannot be broken down into several bilevel problems because we have multiple upper-level variables shared by the lower-level problems. To reduce the number of bilinear terms, we initially only include the two lower-level problems 
corresponding to the two problems that determined the worst case aggregate power flexibility range in step 1). This represents a relaxation of the original optimization problem and by solving it, we obtain an upper bound on the new aggregate power flexibility range and proposed inverter setpoints. Note that this ideal case problem comprises of multiple constraints with bilinear terms. Apart from the dual and strong duality constraints, bilinear terms are also present in the inverter constraints~(\ref{eq:inv_cons_pf}a) and~(\ref{eq:inv_qv}a) for the constant power factor and voltage-reactive power mode, respectively, which make the problem non-convex. Commercial solvers such as Gurobi can handle such constraints efficiently using spatial branching~\cite{gurobi} if the number of bilinear terms are small. We use this approach for solving the ideal case problem. Alternatively, it is possible to employ piecewise McCormick envelopes to relax the bilinear terms~\cite{castro2015tightening}, but we defer this to future work.

\subsubsection{Feasibility check}
The solution obtained by solving the ideal case problem in step 2) may allow for voltage violations at some single-phase nodes in the network since not all follower problems were considered inside the problem. Therefore, we perform a feasibility check where we fix the aggregate power flexibility range and inverter setpoints to the values determined in step 2) and solve all~$4n$ lower-level problems defined in~$\text{P}^\pm$. 
If any of the lower-level problems have maximum or minimum voltage magnitudes outside the limits, we go back to step 2) and add this follower problem to the ideal case problem. 

We iterate between step 2) and 3) in this way until the feasibility check determines that all voltage magnitudes obtained from the~$4n$ lower level problems are within limits. 


\section{Case Study}
\label{sec:sim_res}
In this section, we evaluate our proposed approach using the IEEE 13-node feeder~\cite{schneider2017analytic} and one of the taxonomic distribution feeders from Pacific Northwest National Laboratory (PNNL) \cite{schneider2008modern}. The optimization problem is implemented in Julia and the optimization problem is solved using Gurobi~\cite{gurobi}. All simulations were run on a Windows 10 PC with 3.00 GHz Intel Xeon processor and 16 GB RAM. For our analysis, we solve three optimization problems based on the three inverter reactive power control modes. For the constant power factor mode, we use the inverter power factor~$pf_{\text{G},i}^\phi=0.9$ to define the upper and lower bounds in (\ref{eq:inv_cons_pf}b) and for the constant reactive power mode, we use the power ratio~$\gamma_{\text{G},i}^\phi=0.48$ to define upper and lower bounds in \eqref{eq:inv_cons_q}. The load power factor $pf_{\text{L},i}^\phi$ at node $i$ connected to phase $\phi$ is set to $0.95$.

\subsection{IEEE 13-node Feeder}
Fig.~\ref{13feeder} shows the IEEE 13-node feeder with 75 single-phase solar PV installations at seven nodes, illustrated by house blocks where each block corresponds to aggregation of 5 houses. The maximum apparent power rating of each single-phase solar PV inverter is 60 kVA. We have chosen a current operating point where the PV penetration level, calculated as the ratio of total PV generation (in kW) to the total rated load (in kW), is 45\%. The voltage limits are set to~$\underline{v} = 0.9$ p.u.~and $\overline{v} = 1.1$ p.u. The taps of the voltage regulator connecting nodes 650 and 630 are set to high values and as a result, the system is more prone to overvoltage conditions compared to undervoltage scenarios. So, we will only discuss results which focus on identifying aggregate power flexibility limits~$\Delta{\boldsymbol{p}^-},\Delta{\boldsymbol{p}^+}$ that mitigate overvoltages in the feeder (i.e. the lower-level problem objective is to maximize voltage magnitude). The maximum available aggregate power flexibility in the system is~$\Delta\overline{p}=-\Delta\underline{p}=1.64$ MW.

\setcounter{figure}{3}
\begin{figure}[!t]
	\centering	    
		  \vspace{-.2cm}
	\includegraphics[width=0.30\textwidth]{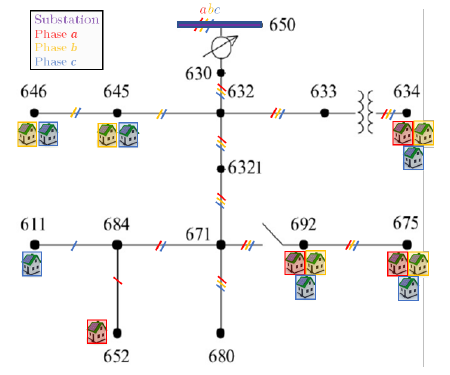}
	    \vspace{-.3cm}
	\caption{Modified IEEE-13 node feeder.}
	\label{13feeder}
	    \vspace{-.5cm}
\end{figure}

\setcounter{figure}{5}
\begin{figure*}[!b]
	\centering	
		  \vspace{-.2cm}
	\includegraphics[width=0.80\textwidth]{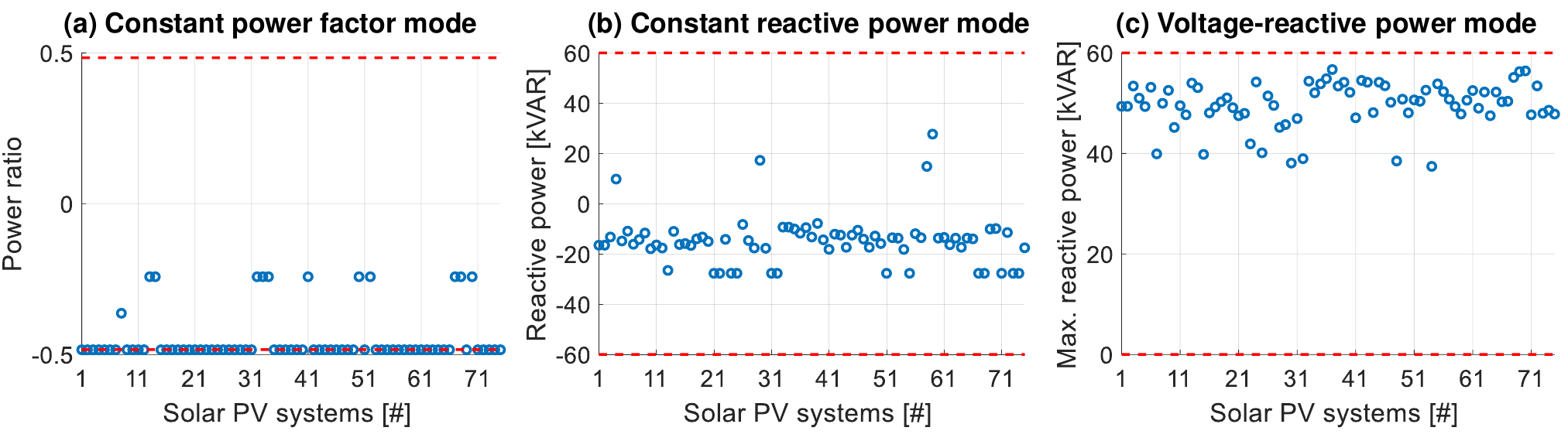}
	    \vspace{-.2cm}
	\caption{Optimal setpoints (blue dots) for different inverter reactive power control modes for the IEEE-13 node feeder. The red dashed lines represent the maximum and minimum limits of the control setpoints.}
	\label{fig:best_case}
	    \vspace{-.5cm}
\end{figure*}

\subsubsection{Worst case aggregate power flexibility limits}
The first step of our iterative solution approach is to solve the worst case problem for each single-phase node separately and identify the location where overvoltages are most likely going to occur. Fig.~\ref{fig:worst_case} shows the aggregate power flexibility limits obtained by solving the worst case problem for the three inverter reactive power control modes. For the constant power factor mode, the lower limit on the aggregate power flexibility~$\Delta \boldsymbol{p}^-$ (blue triangles) is equal to $-1.64$ MW for most of the single-phase nodes, but much closer to zero for some. In contrast, the upper limit~$\Delta \boldsymbol{p}^+$ (orange triangles) is much lower than $1.64$ MW for most single-phase nodes. 
For the constant reactive power mode, the aggregate power flexibility range (i.e. the distance between the blue and orange triangles) is zero for multiple single-phase nodes. 
For the voltage-reactive power mode,
the aggregate power flexibility range at almost all the single-phase nodes is equal to the maximum available aggregate power flexibility. 
From the above results, we see that the choice of reactive power control mode has a very significant impact on the worst-case aggregate power flexibility range.
However, for all three inverter reactive power control modes, the smallest aggregate power flexibility range was obtained when 
maximizing the
voltage magnitude at phase~$b$ of node 675 (single-phase node \#46 in Fig.~\ref{fig:worst_case}). Furthermore, the average computation time to solve the worst case problem with one follower is less than 0.1 seconds.

\subsubsection{Actual aggregate power flexibility limits}
We next solve for the true aggregate power flexibility limits by following the procedure described in Section~\ref{sec:sol_approach}.
The ideal case problem starts with one lower-level problem corresponding to node 675 at phase~$b$, and additional problems are iteratively added following the feasibility check. 
Table~\ref{tab:best_case_res} summarizes the number of iterations and the aggregate power flexibility limits obtained after termination of the iterative process. 

First, we observe that the algorithm only requires a few iterations, regardless of which converter control mode is used. While only one iteration (i.e. only a single follower problem) is required for the problem with constant reactive power mode, the constant power factor and voltage-reactive power modes require that more follower problems are added to the ideal case problem. 
With any of the inverter control modes, the computation time to solve the ideal case problem in the last iteration is less than a second. This indicates that the iterative approach can solve the problem efficiently.

Next, we observe that the aggregate power flexibility limits are same for all inverter reactive power control modes and they are equal to the maximum available aggregate power flexibility in the system. This indicates that if the DSO is able to control the reactive power settings of the inverters, then the amount of flexibility offered by the DERs connected to the grid can be leveraged to the maximum technical limit. 

\begin{table}[!t]
\centering
\caption{Aggregate power flexibility and number of iterations for the IEEE-13 node feeder with different inverter control modes  }
  \vspace{-.2cm}
\begin{adjustbox}{width=0.8\columnwidth}
\begin{tabular}{c|c|c|c}
\toprule
\multirow{2}{*}{\textbf{Inverter mode}} & $\Delta{\boldsymbol{p}^-}$   & $\Delta{\boldsymbol{p}^+}$  & \textbf{No. of}   \\
 &  \textbf{(MW)}   &  \textbf{(MW)} &  \textbf{Iterations} \\
\midrule \midrule
Constant power factor   & -1.64  & 1.64  & 3   \\ \midrule
Constant reactive power    & -1.64  & 1.64  & 1  \\ \midrule
Voltage-reactive power    & -1.64  & 1.64  & 3  \\ 
\bottomrule
\end{tabular}
\end{adjustbox}
  \vspace{-.5cm}
\label{tab:best_case_res}
\end{table}

Fig.~\ref{fig:best_case} illustrates the optimal inverter setpoints for the different inverter control modes. For the constant power factor and constant reactive power modes, the reactive-power injections for most of the solar PV systems are negative (reflected as either a negative power ratio or a negative reactive power injection). This is as expected since the  absorption of the reactive power will lead to lowering of the voltage magnitudes thereby avoiding overvoltage conditions. For the voltage-reactive power mode, the maximum reactive power limit is close to the apparent power limit of the inverters (red dashed line) to ensure that the reactive power absorbed by the solar PV systems is high, which again helps keep the voltage magnitudes within limits.

\subsubsection{Linear approximation accuracy} 
\begin{figure*}[!ht]
	\centering	 
		    \vspace{-.3cm}
	\includegraphics[width=0.80\textwidth]{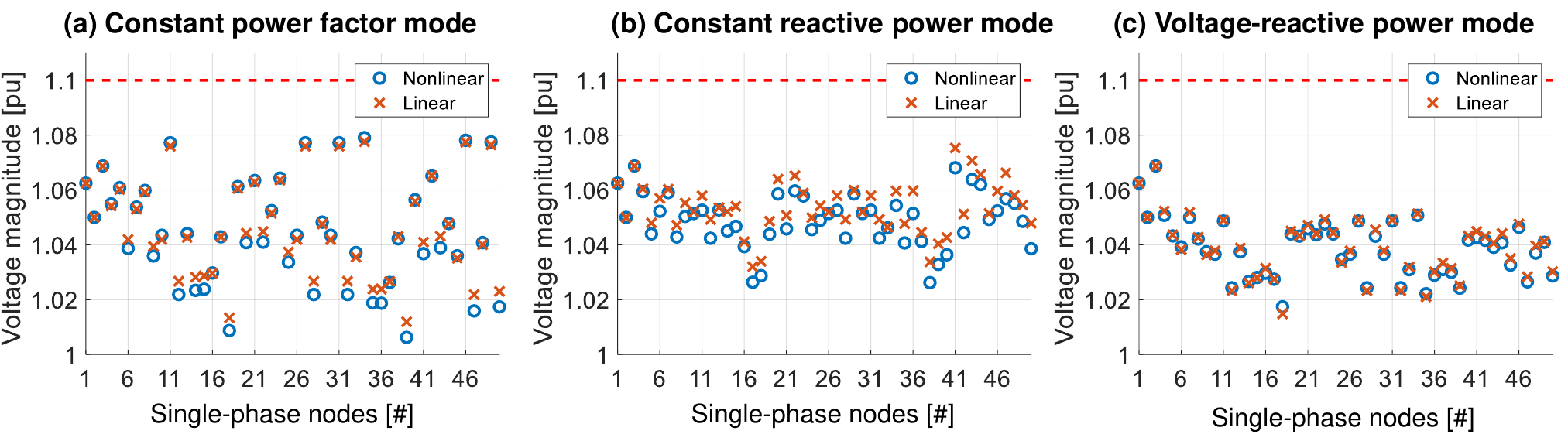}
	    \vspace{-.3cm}
	\caption{Worst-case voltage magnitudes obtained by solving nonlinear (blue circles) and linear (orange crosses) versions of the lower-level problem for IEEE 13-node feeder. The red dashed lines are the upper voltage magnitude limits. From left to right: Results of different inverter control modes. }
	\label{fig:lin_accuracy}
\end{figure*}

To obtain a linear lower-level optimization problem, we linearized multiple constraints. To analyze the accuracy of the inverter setpoints obtained by solving the linearized problem, 
we formulate the nonlinear counterpart of the lower-level problem by making the following changes. (i) We replace the linearized inverter constraints~\eqref{eq:inv_lim_lin} with the original quadratic constraints. (ii) We replace the first order Taylor approximation of the relationship between the voltage variables defined in~\eqref{eq:Vlim_lin} with the original quadratic constraints. (iii) We use the nonlinear power flow equation in rectangular form~\cite{arrillage1983computer} instead of the fixed-point power flow equations~\eqref{eq:FP_rec}.

We then fix the values of the upper level variables (i.e. the aggregate power flexibility limits and inverter setpoints) to the solution identified above and solve the nonlinear subproblem for all single-phase nodes to check if the worst-case voltage magnitudes are within the limits. We also compare the resulting voltage magnitudes with the voltage magnitudes that are obtained from the bilevel problem with the linearized formulation.

\setcounter{figure}{8}
\begin{figure*}[!b]
	\centering	
	  \vspace{-.2cm}
	\includegraphics[width=0.80\textwidth]{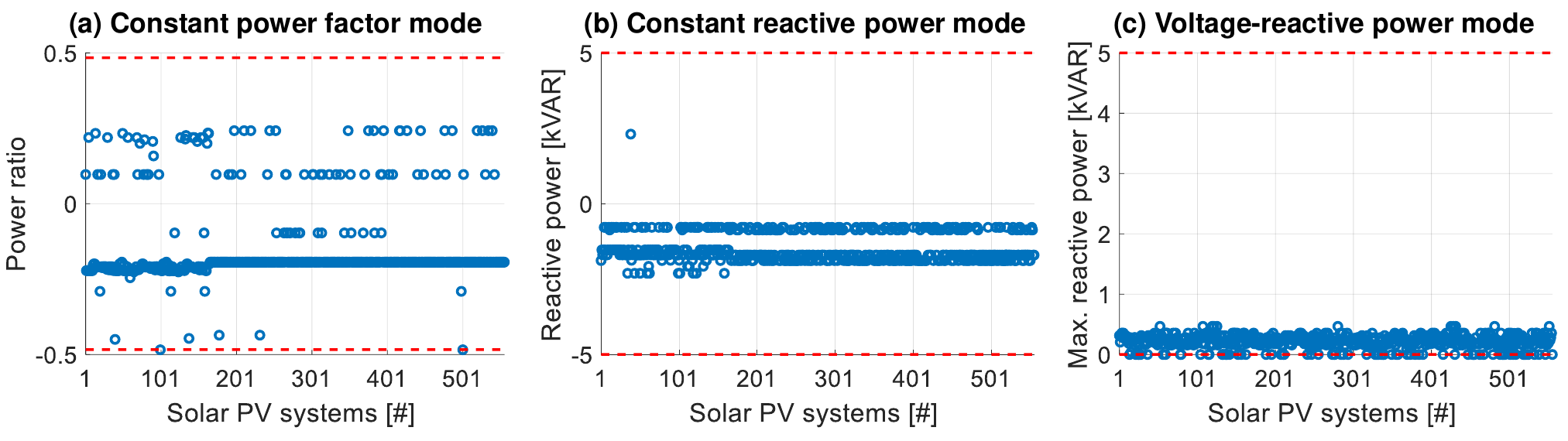}
	    \vspace{-.3cm}
	\caption{R2-12-47-2 feeder results obtained by solving ideal case problem. We show the setpoints determined for different inverter modes. The red dashed lines are maximum and minimum limits on the upper-level variables.}
	\label{fig:best_case_R2}
	    \vspace{-.5cm}
\end{figure*}

Fig.~\ref{fig:lin_accuracy} shows the worst-case voltage magnitudes obtained by solving the nonlinear and linear versions of the problem, for all different inverter control modes. 
Across the three different control modes, we see that the approximation accuracy is high, i.e. the nonlinear voltage magnitudes (blue circles) and linear voltage magnitudes (orange crosses) are close to each other. The maximum error is $\approx$~0.006 p.u., and occurs for the constant reactive power mode. The reason for the higher inaccuracy in this case may be that the solution is further away from the initial operating point which was used to linearize the power flow equations in~\eqref{eq:FP_rec}.
Importantly, 
for all three inverter reactive power control modes, the set-points determined by the linearized problem keep the voltage magnitudes within the limits even we solve the nonlinear problem. 

\setcounter{figure}{6}
\subsection{PNNL Taxonomic Feeder: R2-12-47-2}
\setcounter{figure}{7}
\begin{figure}[!t]
	\centering	   
		    \vspace{-.3cm}
	\includegraphics[width=0.35\textwidth]{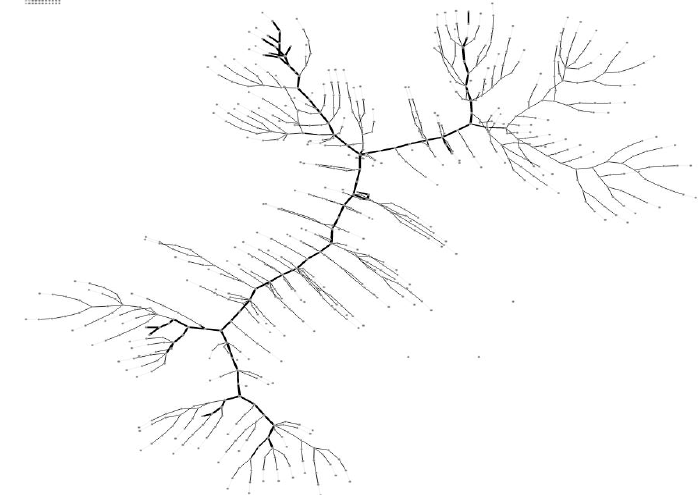}
	    \vspace{-.3cm}
	\caption{Modified R2-12-47-2 taxonomic feeder~\cite{Taxonomicfeeder} visualized using \cite{OMF}.}
	\label{646feeder}
	    \vspace{-.5cm}
\end{figure}

To investigate scalability of our solution method, we next run simulations on the R2-12-47-2 feeder shown in Fig.~\ref{646feeder} which comprises of 820 single-phase nodes~\cite{Taxonomicfeeder}. Solar PV inverters are connected to 556 single-phase nodes in the feeder, and each inverter is rated at 5 KVA to achieve a PV penetration level of 20\% of the total rated load. The voltage limits for this test case are tighter and set to~$\underline{v} = 0.95$ p.u.~and $\overline{v} = 1.05$ p.u. The maximum available aggregate power flexibility before considering voltage constraints is~$\pm4.86$ MW. In this case, we solve the bilevel problem to determine the aggregate power flexibility and inverter setpoints that mitigate both overvoltages and undervoltages in the feeder.

\subsubsection{Worst case aggregate power flexibility limits}
We first 
solve
the worst-case problem for every single-phase node in the network. Different from the IEEE-13 node feeder results, it was observed that all three inverter reactive power control modes provide an aggregate power flexibility range of zero at multiple single-phase nodes in the network. Furthermore, the average computation time to solve the worst-case problem at each single-phase nodes is about 2 seconds. Since there are multiple single-phase nodes with aggregate power flexibility range of zero, we randomly pick one of the them as the first set of lower-level problems to be included in the ideal case problem. 

\subsubsection{Actual aggregate power flexibility limits}
Next, we compare the results obtained after termination of the iterative process where the ideal case problem is solved repeatedly until the feasibility check is passed. Similar to the IEEE 13-node results, the aggregate power flexibility limits determined for all inverter modes are equal to the maximum allowable aggregate power flexibility~$\pm4.86$ MW in the system. It is interesting to note that the feasibility check passed after the first iteration for all three inverter modes. In addition, the maximum computation time to solve the ideal case problem with one follower was approximately one minute. 

The inverter setpoints for the different reactive power modes are shown in Fig.~\ref{fig:best_case_R2}. Compared to the IEEE-13 node feeder results, most of the power ratio setpoints for the constant power factor mode are no longer close to the lower limit. Instead, we see that multiple inverters are injecting reactive power into the grid. This might be because the lower PV penetration level reduces the probability of overvoltages and increases the risk of undervoltages. The inverter setpoints obtained for the reactive power mode are much closer to zero than the lower limit which is similar to the reactive power setpoints obtained for the IEEE-13 node feeder in Fig.~\ref{fig:best_case}(b). For the voltage-reactive power mode, the maximum reactive power setpoints are closer to the lower limit as opposed to the IEEE-13 node results shown in Fig.~\ref{fig:best_case}(c) where the setpoints are near the upper limit. This again indicates that a high reactive power absorption is not required for this test case.
    
    \section{Conclusion}
\label{sec:conc}

This paper explores the coordination between transmission systems, distribution grids and DER aggregators. The main goal is to identify the aggregate power flexibility range that can be provided from distribution-connected resources to the transmission system without causing any constraint violations within the distribution grid itself. We formulate a bilevel optimization task where the upper-level problem represents the decision making of the DSO, which maximizes the aggregate power flexibility and determines DER inverter reactive power setpoints that ensure no voltage violations occur, while the lower-level problem determines the worst-case disaggregation strategy (i.e. assuming that whoever controls the DERs have no information or interest in enforcing internal distribution grid constraints). Our case studies demonstrate that it is important to allow the DSO to choose the DER inverter reactive power setpoints in order to utilize the maximum available flexibility in the system. These setpoints will ensure that the grid is secure even during worst-case conditions. The results for the IEEE 13-bus feeder showed that we are able to obtain high-quality solutions for the inverter setpoints by solving the single-level, strong duality based reformulated problem. Our results on the larger taxonomic feeder demonstrate that the proposed approach can be applied to large, realistic distribution feeders. In both cases, we are able to enable DERs to use their full range of flexibility, even though our formulation assumes worst-case behavior of aggregators.

For future work, we will focus on integrating DERs such as electric vehicles and battery storage systems with temporally-coupled constraints in our optimization framework. We also plan to investigate the impact of limited information about the current operating point on the aggregate flexibility range obtained by the bilevel problem.  

\bibliography{IEEEfull,References}
    
\end{document}